# Trans-generational effect of trained aversive and appetitive experiences in *Drosophila*

Ziv M Williams

Department of Neurosurgery, MGH-HMS center for nervous system repair, Harvard Medical School, Boston MA

Correspondence can be sent to: zwilliams@partners.org

**Associative learning allows animals to rapidly adapt to changes in the environment. Whether and what aspects of such acquired traits may be transmittable across generations remains unclear. Using prolonged olfactory training and subsequent two-forced choice testing in *Drosophila melanogaster*, it is observed that certain aspects of learned behavior were transmitted from parents to offspring. Offspring of parents exposed to distinct odors during both aversive and appetitive conditioning displayed a heightened sensitivity to those same odors. The conditioned responses associated with those odors, however, were not transmitted to the offspring as they displayed a constitutive preference to the parent-exposed stimuli irrespective of whether they were associated with aversive or appetitive training. Moreover, the degree to which the offspring preferred the conditioned stimuli markedly varied from odor-to-odor. These findings suggest that heightened sensitivities to certain salient stimuli in the environment, but not their associated conditioned behaviors, may be transmittable from parents to offspring. Such trans-generational adaptations may influence animal traits over short evolutionary time-scales.**

# Introduction

Jean-Baptiste Lamarck was among first to suggest that certain acquired traits may be heritable from parents to offspring (Cutler, 1925; Koonin and Wolf, 2009). Since then, numerous examples for the effect of parental experience or epigenetic modification on offspring have been observed (Jablonka and Raz, 2009), including phenotypic effects such as changes in coat coloring, reproductive success and longevity (Greer et al., 2011; Holliday, 1987; Kucharski et al., 2008; Rassoulzadegan et al., 2006).

Whether specific *acquired* traits are transmittable from parents to offspring, however, remains less clear. For instance, by repeated conditioning, animals can rapidly learn to associate a sensory cue, such as an odor, with an unconditioned stimulus such as painful shock in order to produce a specific conditioned response such as avoidance (Akalal et al., 2006; Williams and Eskandar, 2006). Several factors, however, are thought to limit the potential transmission of such information across generations. Most notably, learned associations are believed to be held by neural ensembles within the brain which have no obvious way of translating task-specific information to the animals' gametes (i.e., the "Weismann barrier"). Moreover, until recently, it has been believed that information cannot be directly transmitted from proteins or other humoral factors to genetic material. These limitations would, therefore, suggest that no acquired information can be generally transmitted from parents to offspring (Sabour and Scholer, 2012).

The goal of the present study was to (1) determine whether exposure to specific salient sensory stimuli by adults can lead to behavioral effects in their offspring, and (2) determine whether these effects may be behaviorally-specific. While probing these questions using *Drosophila melanogaster*, a remarkable study just came out investigating the trans-generational effect of aversive condoning on mice (Dias and Ressler, 2013). In this study, the authors demonstrated that adult male mice trained on a specific odor paired with an electric shock sire offspring that display a heightened aversive response to that same odor. Moreover, this effect was associated with odor-specific changes in the animal's neuronal circuitry. The aim of this article, therefore, is to present ongoing results from the current study which complement and support their findings. A further aim is to provide additional experiments that may shed light on the mechanism by which parental experiences may affect offspring behavior.

# Methods

*General setup*

Male and female adult flies were cultured together on corn starch medium at 25 degrees Celsius on a standard day-night cycle. Approximately 2-7 days after emergence, the flies were trained together under one of two possible conditions. In the first, they were trained to associate an odor with an aversive stimulus (aversive training) and, in the second, they were trained to associate an odor with an appetitive stimulus (appetitive training).

For the purpose of these experiments, it was important to dissociate learning-specific response after training from a general odor preference/avoidance. Towards this end, flies were trained on only one odor which included either 3-octanol (OCT) or 4-Methylcyclohexanol (MCH). The culture medium was then changed, and larva from flies that underwent 6 days of training were used for testing the next generation. On testing, the flies were given the two odors simultaneously within a T-maze set-up. Under these conditions, the flies could display a preference/avoidance to one or neither odor. Furthermore, comparisons could be made between flies that underwent aversive training, appetitive training or no training. Finally, different groups of flies could be counterbalanced such that the parent flies were exposed to OCT, MCH or neither odor.

*Aversive conditioning protocol*

For aversive conditioning, the flies were placed in 15 cc conical tubes circumferentially covered with an electrifiable copper grid. A mesh was placed on each end of the tube to allow for the free flow of air and odor. Air was first bubbled through distilled water at room temperature and then through a 1:1 mixture of odorant (OCT or MCH) and mineral oil. The air flow would be directed into the fly enclosure through silicone tubing (**Figure 1**, *top*).

For training, the selected odor was pairs with electrical stimulation for 5 minutes for a total of three training sets *per* day. In each set, the flies were first given the odor for 30 seconds. Then, electrical stimulation (100 volts, AC) was given for 1-second durations every 30 seconds for a total of 5 minutes. At the end of the 5 minutes, the odor was removed by gentle suction and the flies were allowed to rest for 5 minutes. This sequence of 5 minutes of stimulation followed by 5 minutes of rest was repeated three times *per* day. Each group of flies underwent 3-5 days of training. Different groups of flies underwent either only OCT training (i.e. they were not exposed to MCH at any time) or MCH training.

*Appetitive conditioning protocol*

A separate group of flies underwent appetitive training. When the flies were 2-7 days old and, over consecutive days, they were first starved for 16-18 hours. After each starvation, they then underwent appetitive training the following day. During appetitive training, they were placed in a 50 cc conical tubes with standard corn meal medium and added sucrose granules. They were allowed to feed on the medium for 6-8 hours while an odor was introduced through the same apparatus described above. The odor was given every hour for durations of 5 minutes each. After feeding on the medium, the flies were removed from the apparatus and starved again. This sequence was repeated for 3-5 days. As above, different groups of flies underwent either only OCT or MCH training.

*Separation and contact between generations*

To limit direct contact between generations, the trained adults (F0 parents) were separated from their larva (F1 offspring). Following training, the adults were plated onto a new culture medium and remained there for 1-2 days. The adults were removed from the culture medium and the larva were allowed to grow.

*Conditioned response T-maze testing*

The aim of the study was to investigate the trans-generational relationship between an unconditioned stimulus (US) such as sucrose or electric shock, a conditioned stimulus (CS) such as the OCT or MCH odors, and the conditioned response (CR) such as odor preference or aversion. To test whether the CR was specific to the CS, two odors were tested simultaneously. Moreover, to test whether the CR was specific to the US, both aversive and appetitive stimuli were tested.

Here, a two-limb T-maze set up was used that presented the flies with the CS in one limb and the other non-conditioned odor in another (**Figure 1**, *bottom*). The concentration of both the conditioned and non-conditioned odors was calibrated such that unexposed wild-type flies would, on average, chose either odor with equal probability. This corresponded to a 1:100 dilution of OCT in mineral oil and a 1:1 dilution of MCH in mineral oil.

During testing, the flies were introduced with suction to the midpoint of the apparatus, and then allowed to freely roam the T-maze for 5 minutes. After this, the air-flow system (similar to the one described above) was connected to each end of the apparatus using silicone tubing. The bottom of the T-maze received low out-flow suction. The flies were allowed to move within the apparatus, as both odors were introduced through each end, for 5 minutes. After the 5 minutes elapsed, the number of flies on each limb was counted. All training and testing was done under dark (low-light) conditions.

*Statistics*

F0 and F1 generations were grown in groups. Each group consisted of approximately 20-60 flies. On T-maze testing the *OCT:MCH preference ratio* was defined as the number of flies counted in one limb of the maze minus the number counted in the other limb divided by the total number of flies in each limb. Flies that displayed no preference (i.e., that remained in the middle of the apparatus after 5 minutes) were not included in this calculation. For convention purposes, a preference ratio of +1 indicates that all counted flies preferred OCT and a preference of -1 indicates that they preferred MCH. A paired, two-tailed Student's t-test was used to determine whether F0 or F1 flies that were trained on a particular odor displayed a significant preference for that odor over the other ($P < 0.05$). To determine whether there was a general effect of training across odors, the preference ratios themselves were compared across the distinct odor-trained groups (unpaired, two-tailed t-test, $P<0.05$).

## Results

Five types of flies were tested. These included flies that (1) received no exposure to the CS or US, (2,3) received an aversive US in association with either OCT or MCH as the

CS and (4,5) received an appetitive US in association with either OCT or MCH as the CS. In most cases, F0 parents were tested after they were removed from the culture medium (i.e. after 3-5 days of training), and their F1 offspring were tested 2-7 days after the pupa hatched. F2 adults were not tested under these settings (see discussion below). A total of 878 F0 flies and 959 F1 flies were included in these experiments and were tested in groups (n). Wild type F0 flies demonstrated a slight but non-significant preference for OCT over MCH with a OCT:MCH ratio of $+0.028\pm0.03$ (n=10; t-test, P=0.73; **Figures 2**). This was similarly true for wild type F0 adults with a ratio of $+0.005\pm0.02$ (n=7; t-test, P=0.92; **Figures 3**).

*Aversive odor conditioning*

F0 adults that underwent aversive training demonstrated a strong avoidance to the trained odor. When trained on OCT, the OCT:MCH preference ratio was $-0.34\pm0.05$ meaning that flies which underwent OCT aversive conditioning preferred MCH and avoided OCT (n=8; t-test, P=0.076). When the flies where trained on MCH, in comparison, the preference ratio was $+0.26\pm0.5$ (n=4; t-test, P=0.42) meaning that they preferred OCT. While individual training results were not significant, the general effect of aversive training on the flies, when considered across both odors, was significant (unpaired t-test, P=0.050).

F1 adults whose parents underwent aversive training continued to demonstrate a differential response to the same odors. However, the response displayed by the F1 offspring was *reversed*. In other words, the F1 adults demonstrated a *preference* towards the CS even though the CS was paired with the aversive US in their F0 parents. Also the degree to which the F1 flies responded to the two odors differed.

Specifically, when F0 adults were trained on MCH, the preference ratio of the F1 offspring was $-0.37\pm0.03$ and significant (n=12; t-test, P=0.021; **Figure 2**). However, when F0 adults were trained on OCT, the preference ratio of the F1 flies was $+0.056\pm0.03$ and was not significant (n=12; t-test, P=0.24). Therefore, the F0 flies who underwent aversive training to MCH demonstrated an aversion to MCH. However, their F1 offspring demonstrated a general preference to MCH. Overall, the effect of F0 training on the F1 flies, when considered across both odors, was significant (t-test, P=0.008).

*Appetitive odor conditioning*

F0 adults that underwent appetitive training had a weaker response to the trained odors when compared to aversive training. When trained on OCT, the preference ratio was $+0.18\pm0.02$ (n=4; t-test, P=0.022) and when trained on MCH, the preference ratio was $-0.14\pm0.02$ (n=4; t-test, P=0.063). In other words, the F0 flies demonstrated a general preference towards the CS. The effect of appetitive training overall was significant (t-test, P=0.001).

F1 adults whose parents underwent appetitive training also demonstrated a slight *preference* towards the same odors but, as before, the degree to which this occurred varied based on which exact odor was trained. Specifically, when F0 flies were trained on MCH, the preference ratio of their F1 offspring was -0.29±0.05 but, strictly speaking, this effect was non-significant (n=6; t-test, P=0.053; **Figure 3**). When F0 flies were trained on OCT, the F1 preference ratio was +0.065±0.04 (n=6; t-test, P=0.46). The general effect of F0 appetitive training on the F1 offspring was significant across odors (t-test, P=0.038).

Therefore, F1 flies whose parents underwent appetitive conditioning demonstrated a slight increase in preference to the same CS but the degree of this response differed between odor types. However, also keep in mind that the absolute *degree* of odor preference is strongly affected by the concentration of the odors themselves. Therefore, for example, increasing the OCT concentration from 1:100 to 1:10 dramatically reduces the preference of the MCH trained flies to MCH (data not shown).

## Discussion

The present findings suggest that exposure to specific sensory stimuli under aversive or appetitive conditions leads to behavioral changes in *Drosophila* offspring. While the findings suggest that the transmission is specific to the CS for certain odors, this does not hold true for the CR. In other words, offspring whose parents underwent prior conditioning did not distinguish between odors that were trained under aversive vs. appetitive conditions. Rather, they demonstrated a non-specific enhanced preference for the salient odor during conditioning by the F0 flies, suggesting that prior parental exposure leads to an enhanced sensitivity to those same odors. The associated conditioned responses, on the other hand, do not appear to be carried over from parent to offspring.

This lack of behavioral response selectivity suggests that the mechanism of transmission is likely not neuronally-mediated and would not, under classical definition, be construed as transmitted "memory" (Carew et al., 1981; Liu et al., 2012). One common theme to the aversive and appetitive conditioning tasks is that they are both known to enhance arousal and other homeostatic systems which can manifest by widespread physiological changes in the flies (Hermans et al., 2011). Whether such changes trigger other odor-specific factors that can be transmitted to the fly's gametes is unclear. Moreover, not all conditioned odors produced the same effect. For example, aversive training with the odor MCH had a significant effect on subsequent offspring behavior whereas OCT did not (or, at least, the effect was weaker). This suggests that the capacity to transmit information about different experienced stimuli differs from odor-to-odor.

Given the above observations, additional studies will be required to shed further light on the mechanism by which parental experiences affects offspring behavior. For example, it may be possible to use dominant temperature-sensitive transgenes, such as UAS-Shi$^{ts}$, in order to selectively and reversibly block specific memory processes (Akalal et al., 2006).

A lack of effect may suggest that the basic mechanism of transmission is not neuronally-mediated and does not directly rely on the formation or reactivation of trained memory representations *per se*. Other tests which may clarify the behavioral effect of parental exposure would include blocking olfaction in the parents, whereby parents are exposed to the odors but specific associations cannot form. In addition, it would be important to test the F0 generation under control conditions whereby the CS is given without a US (e.g. electrical stimulation) or, alternatively, in which the US is given without a CS (e.g. MCH). It would also be interesting to test the effect of simultaneous odor training (i.e. CS+ and CS-) in the parents, rather than in separate groups. Lastly, while flies do not gestate as do mammals and there was no direct exposure of F1 to F0 adults in these experiments, it is conceivable that some of the observed effects were attributable to an indirect induction of early stage larva (i.e., that may have been indirectly exposed to the conditioned and/or unconditioned stimuli). A final caveat in interpreting these preliminary findings is the limited number of groups tested (even though more than 1500 flies in total were used in these experiments). A larger number of groups will be needed to confirm the reproducibility of these experiments.

In sum, the present study provides support for the trans-generational transmission of acquired odor sensitivities in *Drosophila melanogaster*. In particular, these findings suggest a surprising similarity between invertebrates and vertebrate animals (i.e., findings from the present study and those by Dias and Ressler appeared to be made concurrently and without knowledge of the other). The present findings, nonetheless, suggests that it is information about the conditioned stimulus that is principally transmitted between generations rather than the association between the stimulus and a conditioned behavior. Enhancing such sensitivity to a previously salient stimulus may be important in that it could potentially prime subsequent generations to attend to salient or "important" factors in the environment. Such enhanced sensitivity would be valuable under both aversive and appetitive conditions, allowing animals to rapidly hone in on unique cues that were deemed important by prior generations.

# Figures

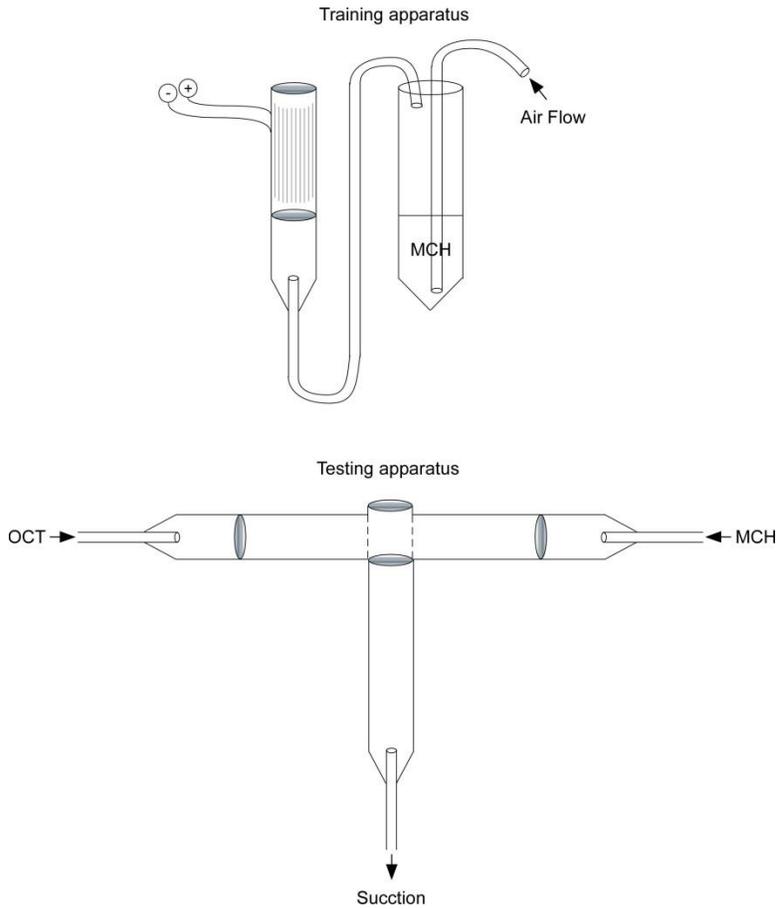

**Figure 1.** Training and testing setup. *Above*, schematic illustration of the training apparatus. Here, flies were placed in an electrified grid that delivered brief 100V stimulation pulses as airflow bubbled in the odorant was introduced into the chamber. *Below*, schematic of the T-maze. Airflow bubbled in odorant was introduced from two ends of the maze.

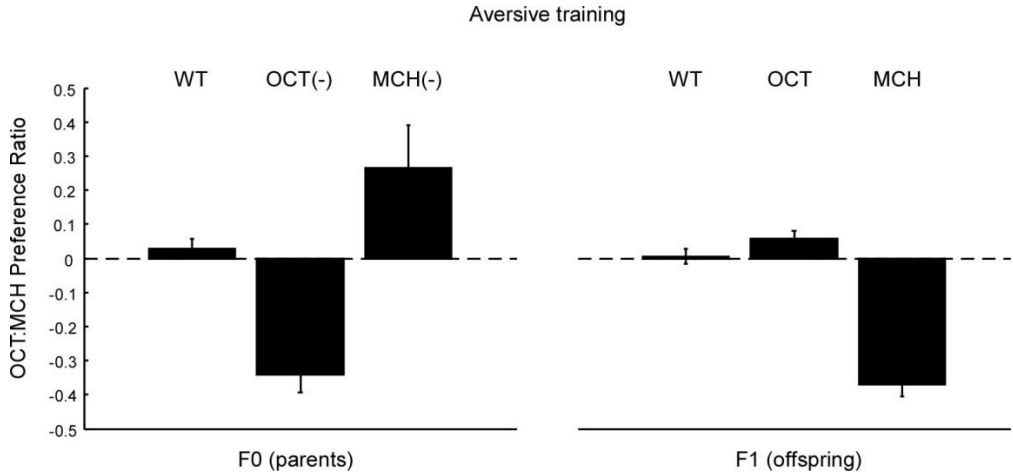

**Figure 2.** Odor preference in F0 and F1 flies following aversive training. The y-axis indicates the preference ration with *positive* values indicating that the flies preferred OCT and *negative* values indicating that the flies preferred MCH. Wild type flies that received no training (WT), F0 flies that were given aversive training with OCT (OCT-), F0 flies that were given aversive training with MCH (MCH-), F1 flies whose parents were given aversive training with OCT (OCT) and F1 flies whose parents were given aversive training with MCH (MCH).

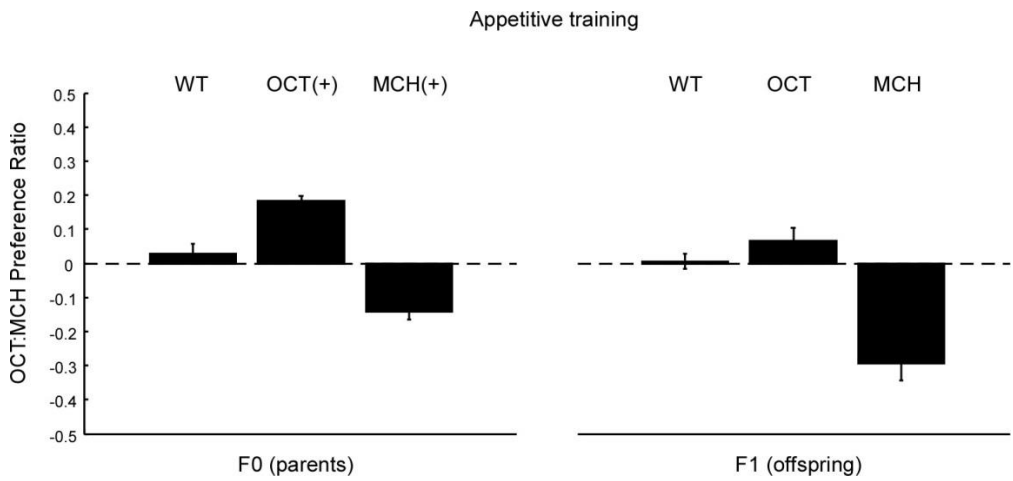

**Figure 3.** Odor preference in F0 and F1 flies following appetitive training. The y-axis indicates the preference ration with *positive* values indicating that the flies preferred OCT and *negative* values indicating that the flies preferred MCH. Wild type flies that received no training (WT), F0 flies that were given appetitive training with OCT (OCT+), F0 flies that were given appetitive training with MCH (MCH+), F1 flies whose parents were given appetitive training with OCT (OCT) and F1 flies whose parents were given appetitive training with MCH (MCH).